\documentstyle{mn}

\newif\ifAMStwofonts

\newcommand{\be}{\begin{equation}}
\newcommand{\ee}{\end{equation}}
\newcommand{\bea}{\begin{eqnarray}}
\newcommand{\eea}{\end{eqnarray}}

\def\L{\Lambda}

\def\r{{\cal R}}

\def\M{{\cal M}}
\def\N{{\cal N}}

\font\cmss=cmss10 \font\cmsss=cmss10 at 7pt
\def\IZ{\relax\ifmmode\mathchoice
{\hbox{\cmss Z\kern-.4em Z}}{\hbox{\cmss Z\kern-.4em Z}}
{\lower.9pt\hbox{\cmsss Z\kern-.4em Z}}
{\lower1.2pt\hbox{\cmsss Z\kern-.4em Z}}\else{\cmss Z\kern-.4em Z}\fi}

\ifoldfss
  \ifCUPmtlplainloaded \else
    \NewTextAlphabet{textbfit} {cmbxti10} {}
    \NewTextAlphabet{textbfss} {cmssbx10} {}
    \NewMathAlphabet{mathbfit} {cmbxti10} {} 
    \NewMathAlphabet{mathbfss} {cmssbx10} {} 
  \fi
  \ifAMStwofonts
    \ifCUPmtlplainloaded \else
      \NewSymbolFont{upmath} {eurm10}
      \NewSymbolFont{AMSa} {msam10}
      \NewMathSymbol{\upi}     {0}{upmath}{19}
      \NewMathSymbol{\umu}     {0}{upmath}{16}
      \NewMathSymbol{\upartial}{0}{upmath}{40}
      \NewMathSymbol{\leqslant}{3}{AMSa}{36}
      \NewMathSymbol{\geqslant}{3}{AMSa}{3E}

       \let\le=\leqslant
       \let\ge=\geqslant
    \fi
  \fi
\fi 

\ifnfssone
  \newmathalphabet{\mathit}
  \addtoversion{normal}{\mathit}{cmr}{m}{it}
  \addtoversion{bold}{\mathit}{cmr}{bx}{it}
  \newmathalphabet{\mathbfit} 
  \addtoversion{normal}{\mathbfit}{cmr}{bx}{it}
  \addtoversion{bold}{\mathbfit}{cmr}{bx}{it}
  \newmathalphabet{\mathbfss} 
  \addtoversion{normal}{\mathbfss}{cmss}{bx}{n}
  \addtoversion{bold}{\mathbfss}{cmss}{bx}{n}
  \ifAMStwofonts
    \ifCUPmtlplainloaded \else
      \UseAMStwoboldmath
      \makeatletter
      \new@mathgroup\upmath@group
      \define@mathgroup\mv@normal\upmath@group{eur}{m}{n}
      \define@mathgroup\mv@bold\upmath@group{eur}{b}{n}
      \edef\UPM{\hexnumber\upmath@group}
      \new@mathgroup\amsa@group
      \define@mathgroup\mv@normal\amsa@group{msa}{m}{n}
      \define@mathgroup\mv@bold\amsa@group{msa}{m}{n}
      \edef\AMSa{\hexnumber\amsa@group}
      \makeatother
      \mathchardef\upi="0\UPM19
      \mathchardef\umu="0\UPM16
      \mathchardef\upartial="0\UPM40
      \mathchardef\leqslant="3\AMSa36
      \mathchardef\geqslant="3\AMSa3E

       \let\le=\leqslant
       \let\ge=\geqslant
    \fi
  \fi
\fi 

\ifnfsstwo
  \DeclareMathAlphabet{\mathbfit}{OT1}{cmr}{bx}{it}
  \SetMathAlphabet\mathbfit{bold}{OT1}{cmr}{bx}{it}
  \DeclareMathAlphabet{\mathbfss}{OT1}{cmss}{bx}{n}
  \SetMathAlphabet\mathbfss{bold}{OT1}{cmss}{bx}{n}
  \ifAMStwofonts
    \ifCUPmtlplainloaded \else
      \DeclareSymbolFont{UPM}{U}{eur}{m}{n}
      \SetSymbolFont{UPM}{bold}{U}{eur}{b}{n}
      \DeclareSymbolFont{AMSa}{U}{msa}{m}{n}
      \DeclareMathSymbol{\upi}{0}{UPM}{"19}
      \DeclareMathSymbol{\umu}{0}{UPM}{"16}
      \DeclareMathSymbol{\upartial}{0}{UPM}{"40}
      \DeclareMathSymbol{\leqslant}{3}{AMSa}{"36}
      \DeclareMathSymbol{\geqslant}{3}{AMSa}{"3E}

       \let\le=\leqslant
       \let\ge=\geqslant
    \fi
  \fi
\fi 

\ifCUPmtlplainloaded \else
  \ifAMStwofonts \else 
    \def\upi{\pi}
    \def\umu{\mu}
    \def\upartial{\partial}
  \fi
\fi

\title{Scaling of voids and fractality in the galaxy distribution}
\author[Jos\'e Gaite and Susanna C. Manrubia]
       {Jos\'e Gaite and Susanna C. Manrubia\\
        Centro de Astrobiolog{\'\i}a, CSIC-INTA, Ctra. de Ajalvir Km.~4,
28850 Torrej\'on de Ardoz, Madrid, Spain.}
\date{Accepted 0000 December 00.
      Received 0000 December 00;
      in original form 0000 October 00}

\pagerange{\pageref{firstpage}--\pageref{lastpage}}
\pubyear{0000}

\begin{document}

\maketitle

\label{firstpage}

\begin{abstract}
We study here, from first principles, what properties of
voids are to be expected in a fractal point distribution and how the
void distribution is related to its morphology. We show this relation
in various examples and apply our results to the distribution of
galaxies. 
If the distribution of galaxies forms a fractal set, then this property 
results in a number of scaling laws to be fulfilled by voids. Consider 
a fractal set of dimension $D$ and its set of voids. If
voids are ordered according to decreasing sizes (largest void 
has rank $R=1$, second largest $R=2$ and so on), then a relation 
between size $\Lambda$ and rank of the
form $\Lambda (R) \propto R^{-z}$ must hold, with $z = d/D$, and
where $d$ is the euclidean
dimension of the space where the fractal is embedded. The physical
restriction $D < d$ means that $z > 1$ in a fractal set. The average size
$\bar \Lambda$ of voids depends on the upper ($\Lambda_u$) and the
lower ($\Lambda_l$) cut-off as 
${\bar \Lambda} \propto \Lambda_u^{1-D/d} \Lambda_l^{D/d}$. 
Current analysis of void sizes in the galaxy distribution 
do not show evidence of a fractal distribution,
but are insufficient to rule it
out. We identify possible shortcomings of current void searching
algorithms, such as changes of shape in voids at different scales or
merging of voids, and propose modifications useful to test fractality in the
galaxy distribution.
\end{abstract}

\begin{keywords}
cosmology: large-scale structure of the universe -- galaxies: clusters:
general -- methods: statistical
\end{keywords}

\section{Introduction}

The morphological properties of the distribution of galaxies are
commonly analyzed by means of the correlation functions of this
distribution, chiefly, the two-point correlation function (Peebles, 1980).
This correlation function is well fitted by a power law up to some
scale (Peebles, 1980). Within this range, the coarse-grained galaxy density
exhibits large fluctuations associated to various structures, namely,
galaxy clusters and superclusters of diverse forms, and voids. Most
studies of galaxy structure have focused on clusters and superclusters,
but the presence of large voids was noted long ago and the size of the
largest voids detected has steadily grown (Einasto, J\~oeveer \& Saar,
1980). The analysis
of voids is a subject of current interest in cosmology (Einasto, Einasto
\& Gramann, 1989; Vogeley, Geller \& Huchra, 1991; El-Ad, Piran \& da
Costa, 1997, M\"uller et al., 2000; Hoyle \& Vogeley, 2002).

On the one hand, the analysis of correlation functions and the
hierarchical structure of clusters and superclusters provides evidence
for a self-similar {\em fractal} structure (Coleman \& Pietronero, 1992;
Sylos Labini, Montuori \& Pietronero, 1998), although
the scale of transition to a homogeneous universe is still matter of
debate (Guzzo, 1997; Wu, Lahav \& Rees, 1999; Chown, 1999; Mart\'{\i}nez,
1999). On the other hand, the scaling properties
of voids are much less studied, but scaling of certain quantities has been
put forward as an indication of self-similarity (Einasto et al., 1989).

Here, we begin by studying the void properties of fractal
distributions in general. Self-similarity is the most obvious property
and is related to the {\em fractal dimension} but there are other
properties worth considering, such as {\em lacunarity} (Mandelbrot, 1977),
which we define in Section 2.
We show in examples how to perform a void analysis to obtain the fractal
dimension and other morphological properties. Then we proceed to compare
with current void analyses of galaxy catalogues, pointing out their
relation with our method and, according to this method, the conclusions that 
can be extracted from these catalogues.

Typically, voids are extracted from galaxy catalogues by using some
{\em void detection algorithm}. These algorithms provide us with a
list of voids, ranked by decreasing size. Therefore, these lists are
suitable for rank-ordering techniques common in statistics 
(Zipf, 1949; Sornette, 2000).
In particular, the {\em Zipf law}, that is, a rank-ordering power law, is
often indicative of fractal behaviour. In our case, a power law cumulative
distribution of voids is expected for a geometric fractal (Mandelbrot, 1977).
This cumulative distribution corresponds to a rank-ordering Zipf law (with
different exponent).

We shall begin studying the rank-ordering of
voids in simple geometric fractals, namely, Cantor sets, where the
Zipf law can be easily proved.  Furthermore, we will consider some
suitable two-dimensional fractals, where the problem of
definition of voids arises. We will extrapolate the results to
three-dimensional fractals. In all cases, we will refer to 
{\em pure fractals}, 
which can be characterized by a single exponent, that is, their fractal 
dimension. We will not discuss the more complex case of 
{\em multifractal sets},
where a whole spectrum of singularity exponents is required to accurately
describe their geometrical properties (Falconer, 1990). 
Finally, we apply the previous results to
current galaxy void catalogues.

\section[]{Zipf's law and fractality in Cantor sets}

Consider a set of quantities $\{\Lambda_k\}$ corresponding to $k=1, 2, \dots$
measures of a phenomenon. It is common-use to measure the probability
distribution $p(\Lambda)$, that is the probability to find an event of
size $\Lambda$, in order to
quantify the statistical properties of the system. An alternative way to
carry out a similar quantification is
provided by the {\em rank-ordering technique}, introduced by
Zipf in the fifties (Zipf, 1949). This procedure highlights the properties of
the large values of $\Lambda$: the largest value is assigned rank $R=1$, the
second largest has $R=2$, and so on. The function $\Lambda(R)$ conveys an
information equivalent to $p(\Lambda)$. In particular, if $\Lambda(R) \propto
R^{-z}$, then $p(\Lambda) \propto \Lambda^{-\alpha}$, with $\alpha = 1+1/z$.
This relation can be explained as follows. Note that $p(\Lambda)$ is the 
fraction of voids with size $\Lambda$. Hence, the total number of voids
with size larger than or equal to $\Lambda$ [which corresponds to the 
function $R(\Lambda)$] is proportional to the accumulated distribution 
$p(\Lambda)$, that is,
\begin{equation}
R (\Lambda) \propto \int_{\Lambda}^{\infty} p(\Lambda) {\rm d} \Lambda \, .
\end{equation} 
If $p (\Lambda) \propto \Lambda^{-\alpha}$, direct integration returns
$R(\Lambda) \propto \Lambda^{1-\alpha}$. Inverting it, we obtain the 
reported relation between $z$ and $\alpha$. 

In order to illustrate the relation between Zipf's law for void sizes and
the geometrical properties of a matter distribution, we begin with Cantor-like
sets defined in the unit interval. If we restrict to deterministic fractals,
a number of relevant quantities can be exactly calculated and clearly
put in correspondence with each other.

A deterministic Cantor set is generated by an iterative procedure.
Its {\em generator} is characterized by three
independent quantities.
First, $r<1$ is the scaling factor. Usually,
the unit interval is divided into $1/r$ pieces of equal length. Of these, $N$
intervals remain for the process to be repeated and ($1/r-N$) are eliminated.
These two quantities completely
define the fractal dimension of the asymptotic set, namely $D=-\log N/\log r$
(Mandelbrot, 1977).
Nonetheless, there are different ways in which the $N$ intervals can be chosen
(in particular, some of them could be adjacent). Therefore,
there is still a degree of freedom
which translates into a variable number $m < N$ of voids (or gaps) in the
generator.
The more adjacent intervals, the fewer gaps and smaller $m$.

The effect of the parameter $m$ in the morphology of the fractal set
is quantified through an appropriate measure of {\it lacunarity},
that is, the quality of having large voids (for a given sample size).
Figure 1 shows three examples of generators and the first iteration for sets
with $N=5$, $r=1/9$ (hence with the same fractal dimension) and different $m$.
The classic triadic Cantor set has $N=2$, $r=1/3$, and $m=1$. Finally, note
that the average length $c(r,N,m) \ge 1$ (in units of $r$) of the $m$ initial
voids can be obtained from the
relation
\begin{equation}
\label{c}
c m + N = \frac{1}{r},
\end{equation}
where the function $c(r,N,m)$ gives an estimation
of the degree of ``adjacency'' of voids of size $r$. In the following,
we write only $c$ for simplicity.

\begin{figure}
\setlength{\unitlength}{1mm}
\begin{picture}(50,70)
\includegraphics{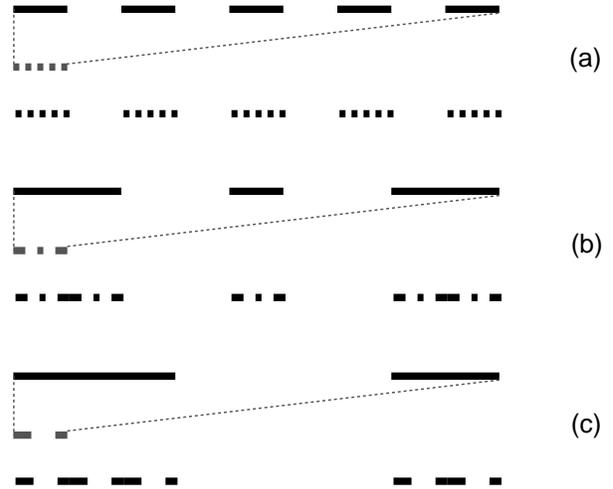}
\end{picture}
\caption{Three generators for Cantor-like fractals in the unit interval and
the first iteration of the algorithm. 
We show the generator and the scaling rule producing the fractal set:
$N$ repetitions of the scaled set are used to construct the next
iteration.
The three examples shown have a variable number of gaps in the generator
(a) $m=4$, (b) $m=2$, (c) $m=1$, other parameters are shared: $N=5$, $r=1/9$.
Hence, these three fractals have the same fractal dimension 
$D=\log 5/\log 9$ but different lacunarity.}
\end{figure}

After the independent quantities $N, \, r,$ and $m$ have been introduced,
we can turn to the explicit calculation of Zipf's law and related
quantities. In the first iteration of the deterministic process, we have $m$
voids with ranks from $R_1=1$ to $m$ and length $\Lambda_1=cr$. In the second
iteration,
there will be $mN$ voids occupying ranks from $R_2=1+m$ to $m+mN$, and their
typical length is $\Lambda_2=cr^2$. In general, in the $i-$th iteration there
are $mN^{i-1}$ voids of average size
\begin{equation}
\label{L}
\Lambda_i = c r^i
\end{equation}
and ranking from
\begin{equation}
\label{R}
R_i= 1 + m \frac{N^{i-1}-1}{N-1}
\end{equation}
to $R_i+mN^{i-1}-1 \equiv R_{i+1}-1$. One can verify that the rank of
the first and the last void in each size class scales in the same way
with the parameters of the system (the function $\Lambda(R)$ is step-shaped
with steps of equal length in logarithmic scale---see the appendix).
For the sake of clarity,
we will use only the value $R_i$ to calculate the explicit form of Zipf's law,
defined in parametric form by Eqs.\ (\ref{L}) and (\ref{R}).
Eliminating the parameter $i$ and arranging terms we get (for large $R$),
\begin{equation}
\label{Zipf}
\Lambda (R) \approx f(r,N,m) \, R^{-1/D} \, ,
\end{equation}
as shown in the appendix, where $D$ is the fractal dimension of the set and
\begin{equation}
\label{f}
f(r,N,m) = \frac{1-rN}{m} \left( \frac{N-1}{m} \right)^{-1/D} \, .
\end{equation}

Mandelbrot (1997) introduced the gaps' length
distribution $N_r$, defined as the cumulative number of gaps with length
larger than a certain given scale $\Lambda$ and proposed that it is
a power law with exponent $-D$ ($N_r \propto \Lambda^{-D}$).
Now notice that the rank
$R_i$ is defined by adding the total number of gaps larger or equal
than the $i$-th gap. Hence, the
gaps' length distribution corresponds to $R(\Lambda)$, which we can get
through inversion of (\ref{Zipf}):
\begin{equation}
R(\Lambda) = F(r,N,m) \, \Lambda^{-D} \, ,
\end{equation}
and where the prefactor
\begin{equation}
\label{lacunarity}
F(r,N,m) = f(r,N,m)^D = \frac{m}{N-1} \left( \frac{1-rN}{m} \right)^D
\end{equation}
is (a measure of) the lacunarity of the fractal set.
Indeed, $F \propto m^{1-D}$ grows with $m$, so a fractal is the more
``lacunar'' the smaller is $F$. $F$ ranges from $(1-rN)^{D}/(N-1) < 1$
for $m=1$ to one for $m=N-1$.
We conclude that $F^{-1}$ is a
measure of lacunarity, in accord with Mandelbrot (1997).

However, there are other measures of lacunarity and, actually,
Mandelbrot concluded that it might be best to consider the fluctuations of
the mass function $\M(\r)$ (Mandelbrot, 1977) (defined as the mean mass 
inside a
ball of radius $\r$ centered on a point, $\M(\r) \propto \r^D$).
This measure of lacunarity is related to the three-point correlation
function and has been the one most employed (see Blumenfeld \& Ball (1993) 
and references therein).

A particular case of the relations above is provided by
fractals with maximal $m=N-1$ (implying $2N-1=1/r$), that is,
with gaps of minimal length. For those fractals, relations
(\ref{Zipf}) and (\ref{f}) hold exactly for all $R$ and the lacunarity
$F^{-1}=1$ (minimal).
The triadic Cantor set and the fractal of Fig.~1(a) are examples of
this particular type.
However, the triadic Cantor set is somewhat special: since $N=2$,
the only possible value of $m$ is one and $F^{-1} = 1$ is its
largest possible value. This explains why the gap is relatively large,
despite the lacunarity being minimal. Nevertheless, we can construct
fractals with its same dimension and larger lacunarity, for example,
by taking $r=1/9$ and $N=4$; namely, the cases $m=1,2$. The case
$m=3$ gives rise to fractals with $F^{-1}=1$, one of which
actually is the triadic Cantor set.

\section{Scaling of voids in dimensions 2 and 3}

The definition of void is simple and clear-cut in dimension $d=1$,
because a point divides a segment into two disconnected parts. When dealing
with point-sets in higher dimensions, voids are usually ill defined,
since empty areas or volumes are (usually) connected.
Indeed, the factor $c(r,N,m)$ in (\ref{L})
was taking care of the connection between adjacent voids, and for $d > 1$ more
than one definition is possible. A possible generalization of the definition
in the previous section for $d>1$ would be that only voids of equivalent size
(that is to say, in a given iteration) are allowed to coalesce into a larger
void. In this case, our results can be straightforwardly generalized
and Eq.~(\ref{Zipf}) reads

\begin{equation}
\Lambda (R) \approx \frac{1-r^d N}{m} \left( \frac{N-1}{m} \right)^{-d/D}
R^{-d/D} \, ,
\label{Zipf-d}
\end{equation}
where now $\Lambda$ stands for the area or the volume (in units of $1/r^d$) in
dimension $d=2, \, 3$, respectively.
A particular generator is the one that starts with a square and
removes a number of the $1/r^2$ square parts in a manner symmetrical
with respect to a diagonal. The result is just a cartesian product of
one-dimensional Cantor sets. The simplest example is obtained by
taking $r =1/3$ and removing five patches forming a cross out of the nine 
initial ones. The fractal so generated is the cartesian product of triadic
Cantor sets. Clearly, the complementary set of these fractals is connected,
but one can {\em assume} that the voids produced at every iteration are
independent; then, one gets Eq.~(\ref{Zipf-d}).

Other definitions of what constitutes a void are also possible. If we change
our definition, the prefactor of $\Lambda (R)$ will change its precise form,
and the estimated lacunarity of the fractal will also change accordingly.
Nonetheless, the scaling form of the Zipf law for void sizes remains 
unchanged, since it is independent of $m$ and $c$.
In the following, we show through numerical examples that any reasonable
definition of void which is coherently applied at all scales returns the
correct scaling for the Zipf law, and thus allows one to obtain quantitative
information on the geometry of the distribution of points in the fractal set.

\subsection{Voids in random fractals}

In order to move towards the description of
fractals arising in natural processes, we should
first relax the deterministic character of their construction.
Consider now a
two-dimensional set for which $r=1/3$ and $N=4$, but where the five areas
to be removed at each step in the construction are chosen at random.
However, to mimic the observed morphology of the galaxy distribution,
we must impose some constraints:
the galaxy distribution has been characterized as a sponge-like network,
with filaments and walls where galaxies accumulate (Gott, Melott \&
Dickinson, 1986).
In two dimensions, we should generate a fractal with some trace of
filamentary structure. A rough way to achieve this is by constraining
the five patches removed at every step to form a particular ``convex-like'' 
shape, namely, a four-piece square with an extra piece adjacent to one side.

Randomization worsens the scaling range, so one should iterate the
random generator many more times than the deterministic one to keep the
scaling range similar. Alternatively, we will choose to
average over independent realizations of the fractal constructed with
the same number of iterations to improve the measures.

Since the problem to define voids in two-dimensional sets is similar to
the one arising
in three dimensions, we will work with $d=2$ and extrapolate
our results to higher dimensionality. Fig.~2 represents a fractal constructed
by removing five randomly chosen (connected) patches of area $r^{2i}$ at
iteration $i$. Now, the possibility arises that voids produced in subsequent
iterations are adjacent to previously existing ones and result in a more or
less apparent increase in the
size of a large void. As we pointed out, we wish to design an algorithm to
find the voids in our set and apply it at all scales. Our working hypothesis
is that the precise shape of the void is not relevant to recover the scaling
behaviour of Zipf's law, as long as it is kept constant through iterations, 
and thus it is not relevant either to recover the fractal dimension 
of the associated distribution of matter.

\begin{figure}
\setlength{\unitlength}{1mm}
\begin{picture}(50,50)
\includegraphics{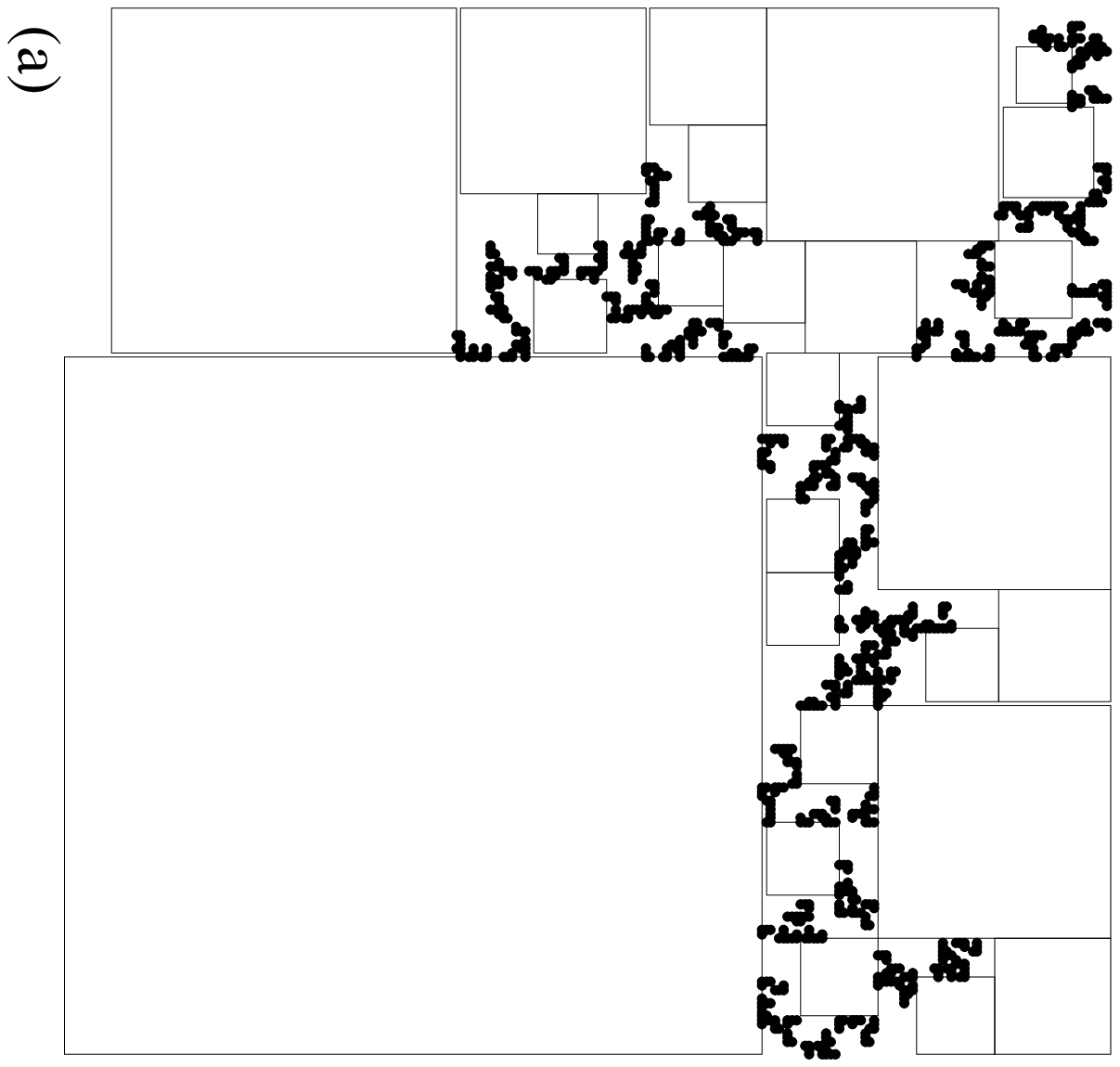}
\end{picture}
\begin{picture}(50,50)
\includegraphics{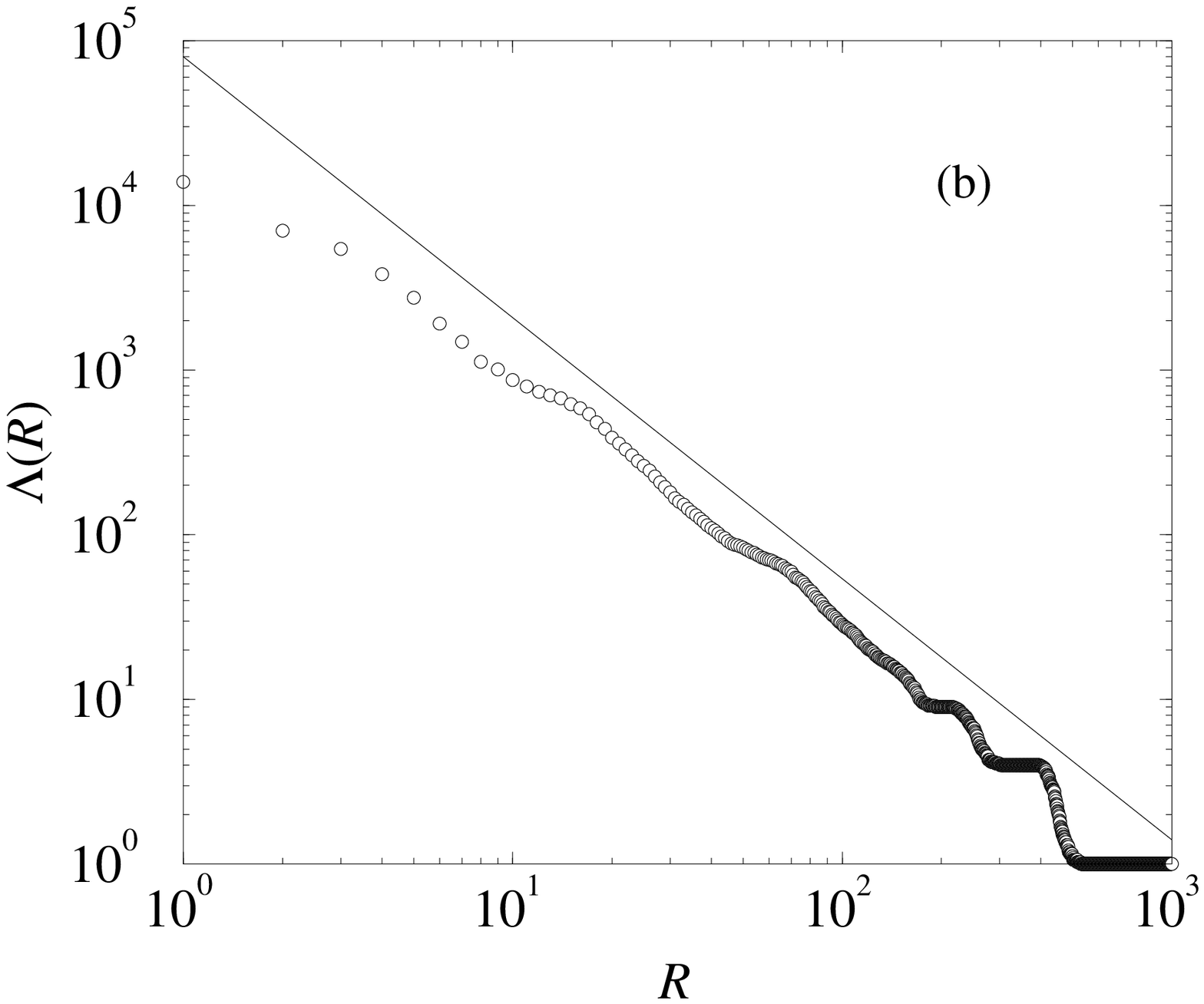}
\end{picture}
\vspace{100pt}
\caption{(a) Fractal point set generated by randomly removing
$5$ connected intervals of linear size $r=(1/3)^i$ at each iteration $i$.
The squares correspond to the largest 25 voids found in the system through the
algorithm described in the text. (b) Zipf law for void sizes calculated
according to the same algorithm. The solid line has slope $-2/D$, where
$D=\log 5/\log 3$. The normalization of this line (its height at the origin) 
is arbitrary. The numerical data has been averaged over 500 independently
generated fractals.}
\end{figure}

Simple recursion relations of the previous type (Section 2) cannot be inferred
when studying
fractal sets arising from physical processes. Instead, one faces a set of
points irregularly distributed in space and has to resort to other methods to
estimate the distribution of voids. One of the simplest ways of defining
an area devoid of points in the structure is the following:

\begin{enumerate}
\item Coarse-grain your system by defining elementary cells such that
there is at the most one point per cell (and defining so a lower
cutoff to scaling, see section 4).
\item Decide for regular elements to cover empty areas (say a square
or a circle in $d=2$).
 \item Locate the largest square/circle centered at each empty cell
(its boundary is limited by filled cells or sample boundaries).
\item Select the largest one, which is by definition a void of size
$\Lambda_1$, and assign it rank $R=1$.
\item Fill the cells in the selected void (equivalent to those in
the fractal set).
\item Repeat the procedure with the remaining empty cells until all of them
are covered (this is somehow reminiscent of the box-counting method
to estimate the fractal dimension (Falconer, 1990)).
\end{enumerate}
When the algorithm finishes, an ordered list of voids of decreasing size
is produced.
In Fig.~2a we represent an example of the first stages of the
void-finding algorithm applied to a random fractal
generated with the ``filamentary''
generator, while Fig.~2b represents the obtained
function $\Lambda (R)$ for square voids. As can be seen, the
voids measured with the previous algorithm follow indeed the scaling expected
according to the analytic prediction for deterministic fractals. We have
applied our method to fractals constructed with several different
generators in $d=2$ and, in all cases, have
obtained a good quantitative agreement between the predicted slope $-d/D$
and the numerically obtained one. 

\begin{figure}
\begin{picture}(50,50)
\includegraphics{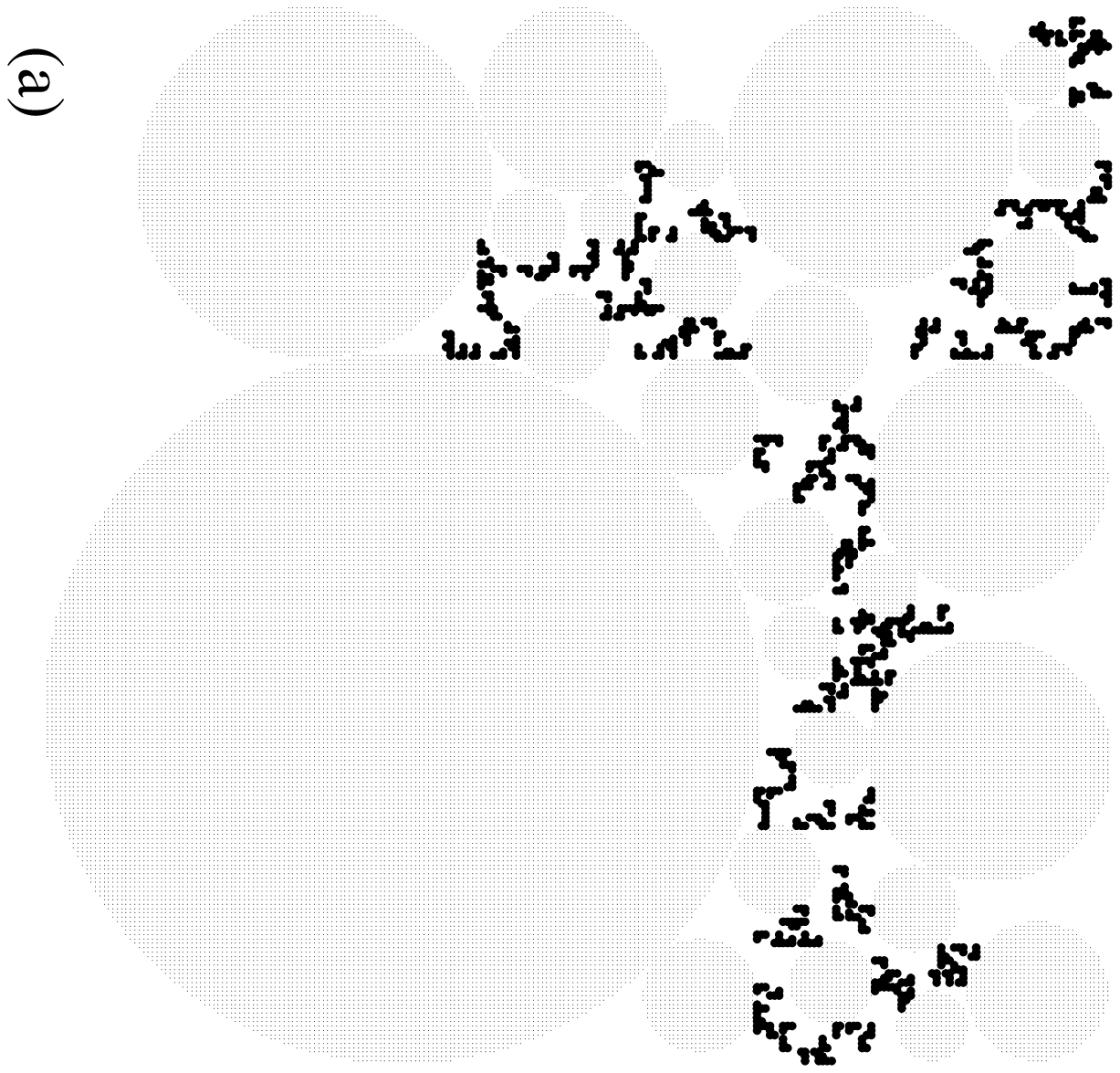}
\end{picture}
\begin{picture}(50,50)
\includegraphics{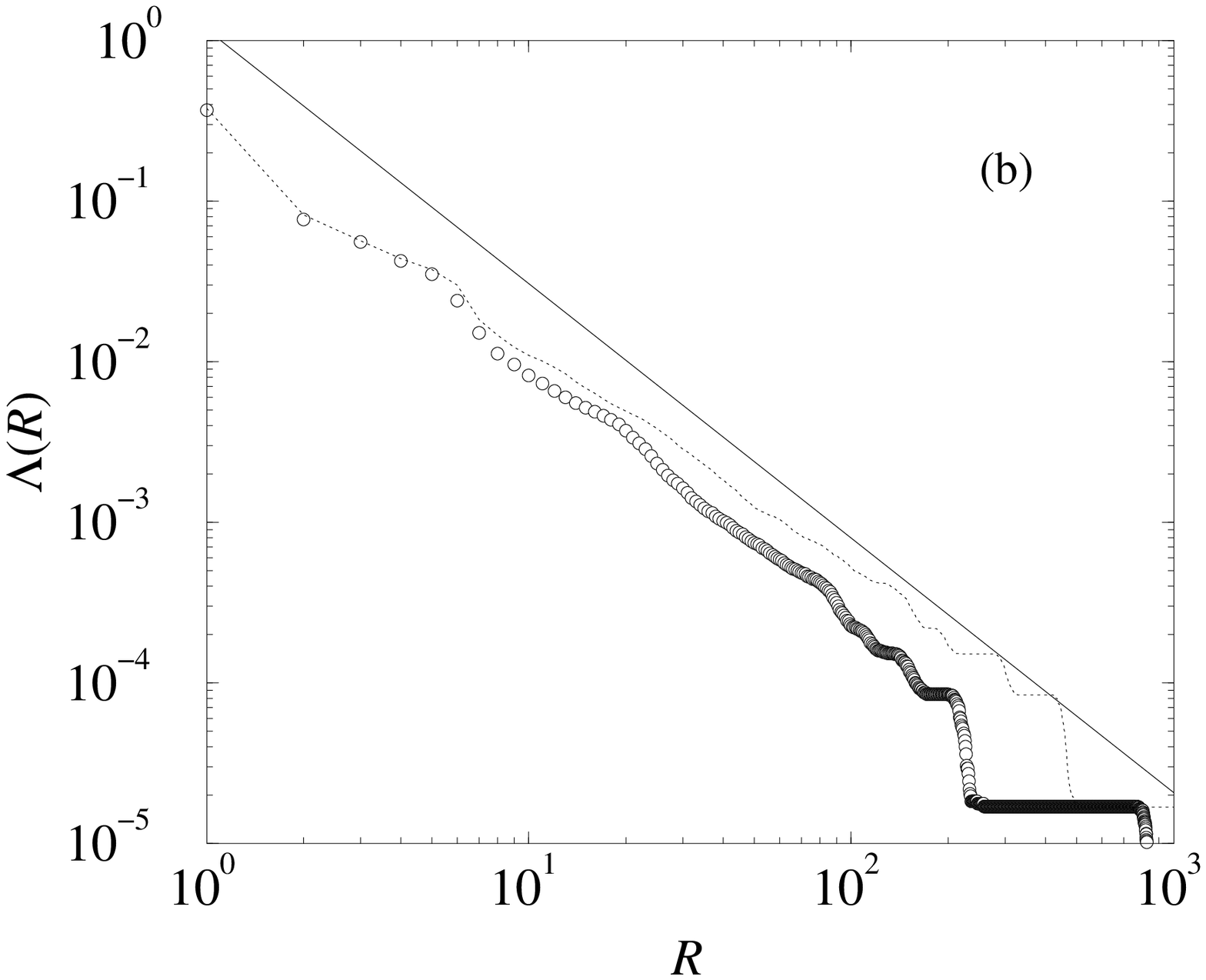}
\end{picture}
\vspace{350pt}
\caption{(a) The 25 largest circular voids found in the same set of Fig.~2
taking into account finite-size corrections. (b) Zipf's law. The dashed line
shows $\Lambda (R)$ for circular voids calculated according to the algorithm
described in section 3. The circles show $\Lambda (R)$ for circular voids 
when only those with boundaries limited
by the fractal set are counted. The scaling improves in the second
case. Again, the solid line only indicates the expected slope $-2/D$. 
Numerical results have been averaged over 100 independently generated 
fractals.}
\end{figure}

Another regular shape for voids that we have investigated is the circular one.
Although largest voids clearly become even larger when a circular 
coverage is used, boundaries among voids define smaller voids which, however,
are not limited by points in the fractal set (see Fig.~3). 
A first attemp to correct for these boundary effects
would be to reject, in the final count,
the voids that do not touch at least one point of the fractal set (that is,
that only touch the external boundaries and/or other voids). This 
modification of our previously
defined algorithm returns a better scaling for average sizes in the case
of circular voids (both far from
the boundary and from the elementary cell, see Fig.~3). These size classes
were overloaded with ``artificial'' voids placed among previously defined
voids and/or external boundaries.

\section{Mean size of voids in a fractal}

As shown before, the distribution of voids in a fractal is a power law
and, therefore, the mean size of voids is not a characteristic value,
being dependent on the upper and lower cutoffs to the fractal
scaling. Actually, the mean size of voids has been employed as a test
for fractality, under the assumption that in a fractal the mean size
of voids is proportional to the size of the sample, that is, the
upper cutoff (Einasto et al., 1989). We shall show that this is not always the
case: in fact, the mean size of voids is usually dependent on both
cutoffs.

To calculate the mean size of voids we must use the probability
distribution of sizes $p(\Lambda) \propto \Lambda^{-\alpha}$, with
$\alpha = 1+D/d$. Then,
\begin{equation}
\label{integral}
{\bar\Lambda} =\frac{\int_{\L_l}^{\L_u} \L\, p(\L)\, d\L}
{\int_{\L_l}^{\L_u} p(\L)\, d\L},
\end{equation}
where $\L_l$ and $\L_u$ are the lower and upper cutoffs, respectively.
The computation of the integrals is straightforward and
\begin{equation}
\label{LUC0}
{\bar\Lambda} = \frac{-\alpha+1}{-\alpha+2}\,
\frac{\L_u^{-\alpha+2}-\L_l^{-\alpha+2}}{\L_u^{-\alpha+1}-\L_l^{-\alpha+1}}.
\end{equation}
For large ratio $\L_u/\L_l$ (as required to have a reasonable scaling range)
and taking into account that $1 < \alpha < 2$,
\begin{equation}
{\bar\Lambda} \approx \frac{\alpha-1}{2-\alpha}\,
\L_u^{2-\alpha}\,\L_l^{\alpha-1}.
\label{LUC}
\end{equation}
This expression cannot be reduced anymore and depends on both cutoffs.
Note that if $D=d/2$ then $\alpha-1 = 2-\alpha = 1/2$ so that the
mean size is just the geometric mean of both cutoffs (so to speak,
both contribute equally); if $D>d/2$ then $\alpha-1 > 2-\alpha$ and
the lower cutoff contributes more to ${\bar\Lambda}$, and viceversa if
$D<d/2$.

We now discuss
what values we should take for $\L_u$ and $\L_l$
in point distributions: whereas $\L_u$ is the well determined
size of the sample, the lower cutoff
is trickier. Since a random fractal has a random component superposed on
the deterministic algorithm that generates it, on the lowest scales the
random component dominates and the distribution is approximately
of Poisson type but with a very low mean number of points in the associated
volume ({\em shot noise}).
The crossover scale from the Poisson to a correlated (fractal) regime
such that the number function%
\footnote{The number function is the mean number of particles inside a
ball of radius $\r$ centered on a particle and equals the mass function
$\M(\r)$ divided by the mass of a particle.}
${\N}(\r) = B \r^{D}$ is $B^{-1/D}$ (this scale has been defined by 
Balian \& Schaeffer (1989) in a more general context).
It is not totally clear what value is adequate for $\L_l$ in a galaxy
catalogue.
It must be larger than (0.1 Mpc)$^3$ but it can be considerably larger,
according to the way the catalogue has been compiled and the algorithm
selected to find voids. At any rate, keeping $\L_l$ fixed one obtains
that ${\bar\Lambda}$ is not proportional to $\L_u$ but rather to a
power of it with exponent $2-\alpha = 1-D/d$ such that $0 < 1-D/d <1$
($d=3$). A value of this exponent close to 1 (as reported by Einasto et al. 
(1989), see also Section~5)) would imply a fractal
dimension $D \ll 1$. A set with $D = 0$ is not really a fractal, but a 
collection of isolated points. 

We have carried out numerical measurements on 2-dimensional fractals of
known dimension $D$ to test the accuracy to which a real fractal sample
follows the scaling (\ref{LUC}). Figure~4 depicts some of our results.
There, we have generated a fractal with $N=3$ and similarity ratio $r=1/2$,
hence $D=1.58$. The single patch to be removed at each iteration was chosen
at random (see insert in Fig.~4).
In order to see how $\bar \Lambda$ depends on $\Lambda_u$ we
have kept the lower cut-off fixed and equal to the size of the individual cell,
$\Lambda_l=1$ and varied $\Lambda_u$. As long as $\Lambda_l \ll \Lambda_u$
we observe a neat scaling with the predicted exponent $2-\alpha$. Next, the
upper cut-off was kept fixed to its maximum value for this fractal set, 
$\Lambda_u = 2^{14}$. The
variation of $\Lambda_l$ was carried out in practice by averaging only
over voids of area equal to or larger than $\Lambda_l$. In this second case, 
the asymptotic scaling exponent is $\alpha-1$. Numerical results are compared 
with the two curves obtained from Eq.~(\ref{LUC0}) with $\Lambda_l=1$ and 
$\Lambda_u=2^{14}$, respectively. Only when both cut-offs become of comparable 
order are deviations from scaling observed.

\begin{figure}
\begin{picture}(50,80)
\includegraphics{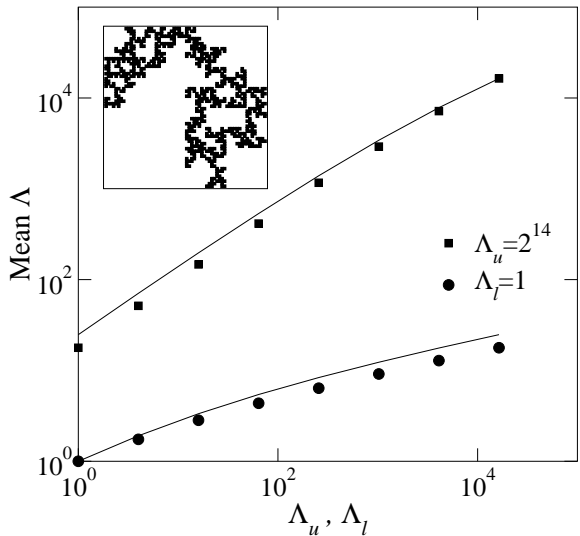}
\end{picture}
\vspace{150pt}
\caption{Scaling of the mean size of voids when one cut-off is varied and
the other kept fixed. The insert depicts a random fractal 
with $N=3$ and $r=1/2$ embedded in a space of dimension $d=2$. 
Solid lines correspond to a numerical solution of the
expression for ${\bar\Lambda}$ (\ref{LUC0}) corresponding to this fractal 
(with $D=1.58$).
The slight difference between numerical and analytical results is due to the
substitution of a sum (over discrete void areas) by an integral in 
the expression for ${\bar\Lambda}$ (\ref{integral}).}
\end{figure}

\section{Galactic voids}

Galactic voids are vast regions of space apparently devoid of luminous matter
(galaxies). Several authors have developed algorithms in order to detect
the extent and frequency of such regions in current galaxy catalogues
(Einasto et al., 1989; Kauffmann \& Fairall, 1991; Hoyle \& Vogeley, 2002).
The aim
of these studies is to gain a better understanding of the morphology of
the distribution of galaxies, in order to, eventually,
correlate it with the physical mechanisms
responsible for the observed structure. A first step in this direction has
been the comparison of measures of voids 
made on the current galaxy catalogues with measures on
$N$-body simulations of cold dark matter (M\"uller et al., 2000; 
Arbabi-Bigdoli \& M\"uller, 2002). The problem of
defining what constitutes a void in $d=3$ has been ubiquitous, and all these
studies have solved this indeterminacy in different ways. Interestingly, all
of them have looked for maximal volumes of approximately convex shape (but
differently shaped depending on the area studied) inscribed among matter
points, while none has considered regular volumes. Our main point
here is that, when trying to recover quantitative information, shape matters,
implying that voids have to maintain their shape at all scales. On the one
hand, we have shown that this criterium permits us to obtain information on a
fractal distribution of matter. On the other hand, our definition eliminates
certain arbitrariness in void finding algorithms, for instance the amount
of overlap between voids to be merged into a single larger void.

With these caveats in mind, we have examined two studies reporting large
voids and examined the function $\Lambda (R)$ that they produce. Recently,
Hoyle \& Vogeley (2002) have examined the Point Source Catalogue
Survey (PSCz) and the Updated Zwicky Catalog (UZC) for the presence of voids.
The volume of the 35 and 19 largest voids (respectively) is plotted in Fig.~5
attending to their rank. Although, in these cases, $\Lambda (R)$ is relatively
well fitted by a straight line for the largest voids, the slope is too low to
represent the complementary set (that is, the set of voids) 
of a fractal distribution of matter.
There is a physical restriction to the exponent of the scaling law, since the
dimension of the fractal cannot be larger than that of the embedding space,
that is to say, $d > D$. This implies that the exponent $-d/D$ has to be larger
than unity in absolute value. We
represent in Fig.~5 this restriction and observe that the putative slopes are
much lower than this value.
The classical value of the fractal dimension, deduced from the two-point
correlation function $\xi(r) \propto r^{-\gamma}$, is $D = 3 - \gamma
\simeq 3 -1.8 = 1.2$. However, recent reanalyses of the galaxy catalogues
(Sylos Labini et al., 1998) yield a larger value, namely, $D \simeq 3 -1.1 
= 1.9$.
In this case, the expected exponent for $\Lambda (R)$ {\it vs} $R$ would be
about $-1.5$, which we represent as a dotted line in Fig.~5. 

Ten years ago, Kauffmann \& Fairall (1991) developed an algorithm to search 
for voids. They reported a list of 129 'significant' voids
obtained from the merged Southern Redshifts Catalogue and the Catalogue of
Radial Velocities of Galaxies. Their list is represented in Fig.~5 together
with the previous data. Apart from an initial almost flat stage,
we have found that this
function is reasonably well fitted by an exponential law (not shown
in the figure).
This does not correspond to a fractal geometry and would rather correspond
to a Poisson distribution of points, where voids much larger than
the volume per point should be exponentially suppressed. However,
they define a {\it significant void} as one that ``\dots occurs in the random
catalogue simulations with probability less than 1 per cent'' (see Kauffmann
\& Fairall (1991) for more details).
Moreover, they take as the reference Poisson distribution for a given
catalogue the one with the same number of points but randomly distributed.
While one certainly cannot consider voids as significant below the scale
of the lower cutoff $\Lambda_l$, where the distribution can be effectively
considered Poissonian (as remarked in section 4), their procedure produces
a Poissonian distribution with a much larger mean interparticle distance.
Hence, even though very large voids are almost
always significant in their sense, this is not the case for average- and
small-size voids, which occur frequently in a random catalogue with
the same number of points. Those voids are
not included in the list they provide, and hence the distribution is strongly
depleted in the mean and small-void domains.

\begin{figure}
\begin{picture}(50,50)
\includegraphics{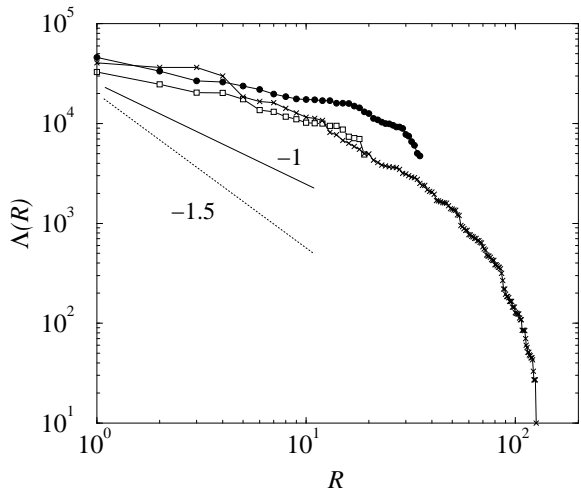}
\end{picture}
\vspace{140pt}
\caption{Zipf's plot for three available void catalogues. Data represented as
circles (PSCz) and squares (UZC) are from Hoyle \& Vogeley (2002), units are
h$^{-3}$Mpc$^3$; data in
crosses from Kauffmann \& Fairall (1991), units are Km$^3$s$^{-3}$. Under the 
hypothesis that
matter is self-similarly distributed in the universe, the scaling expected 
for void volumes is
$\Lambda (R) \sim R^{-3/D}$, where $D$ stands for the fractal dimension
of the galaxy distribution. The solid line has slope $-1$ (indeed, any function
$\Lambda (R)$ has to have slope larger than unity in absolute value). The
dotted line signals the expected scaling if, according to recent measures,
$D \simeq 2$, leading to the $-1.5$ scaling relation shown in the figure. The 
dotted and the solid lines just indicate the (expected) slope of 
$\Lambda (R)$ in each case.}
\end{figure}

Einasto and co-workers (1989) have shown that mean void diameters
increase with the sample size in a power-law manner. They have identified
this fact as indicative of self-similarity in the matter distribution. Their 
qualitative results agree with the calculations here reported (section~4).
However, the quantitative result does not agree with our prediction. In the 
work by Einasto et al. (1989), we understand that measurements were carried 
out in such a way 
that the lower cut-off was kept fixed, while the upper cut-off varied. One
expects then a scaling of the form ${\bar \Lambda} \propto 
\Lambda_u^{\beta_u}$, with $\beta_u = 1-D/d $. Since $0 \le D \le d$, the
exponent is bounded, $0 \le \beta_u \le 1$, with $\beta_u=0$ corresponding 
to a homogeneous distribution of points and $\beta_u=1$ to an (almost) empty
set of points. This second value is the one reported by Einasto et al. (1989).
 
All our considerations apply to galaxies as particles with no
features, in other words, ignoring type and luminosity. It has been
argued that voids could be populated by faint galaxies, called {\em field
galaxies} to distinguish them from ``normal'' cluster galaxies 
(El-Ad et al., 1997; Hoyle \& Vogeley, 2002).
Luminosity can be taken into account by generalizing 
the concept of a fractal distribution to a multifractal distribution 
(Sylos Labini et al., 1998) which we do not consider here. If the neglected 
field galaxies have a distribution with geometrical properties different 
from the more luminous ones, 
the multifractal model would not apply. However, the standard biased galaxy 
formation picture attributes similar scaling properties to the distributions 
at various biasings (Gabrielli, Sylos Labini \& Durrer, 2000), in accord
with a multifractal distribution. A sort of biasing could be mimicked for a 
pure fractal by randomly removing a fraction of points, which would not alter 
its scaling properties. At any rate, a thorough analysis of this question 
falls beyond the scope of this work. 

\subsection{Sources of deviation from scaling}

Apart from our previous discussion on the way in which voids are defined and
counted, there are a number of mechanisms which, in our understanding, could
produce a significant deviation from the scaling regime. We list them and
briefly discuss their effects. Sometimes the source of deviations can be
identified and even eliminated. But often, even in what sense they would
affect Zipf's plot is unclear. The following list might not be exhaustive, but
some or even all of the listed problems can affect the observations to date.
However, note that scaling corresponding to $D > d$ should not occur, since it
is physically forbidden.

\begin{itemize}

\item Finite size effects -- Usually, it is unavoidable to use a ``boundary''
to limit the fractal set that we are measuring. We have already seen that, in
particular for certain shapes of voids, systematic deviations from scaling
can be obtained. 
\item Scale-dependent dimension of galaxy distribution -- It has been
recently reported (Bak \& Chen, 2001) that the dimension of the galaxy 
distribution varies with the observation scale. It grows smoothly from zero 
when approaching the size of single galaxies to three at the scale where the
transition to homogeneity%
\footnote{The concept of homogeneity involves some subtleties
(Gaite, Dom\'{\i}nguez \& P\'erez-Mercader, 1999). Here and henceforth, we 
mean by homogeneity that the relative density fluctuations are small.}
takes place. Since voids
involved in Zipf's plot would cover the whole range of sizes, there
might be some systematic deviations in case the fractal dimension is
scale dependent: $D$ decreases with decreasing scale, hence the exponent of
the rank-ordering plot increases in absolute value, and the function becomes
concave from below.
\item Transition to homogeneity -- There cannot be voids larger than
the characteristic length at which the universe becomes homogeneous;
but the characteristic size of the largest voids is an independent
scale (Balian \& Schaeffer, 1989) that could be significantly smaller than the
homogeneity scale. Furthermore, the breakdown of scaling in the void 
distribution at the
characteristic size of the largest voids might suggest that the recent
observations reported by Hoyle \& Vogeley (2002) and returning a flat 
$\Lambda (R)$ are related to it.
\end{itemize}

\section{Conclusions}

There is a quantitative and well-defined relationship between the dimension
$D$ of a fractal set and the exponent of Zipf's plot $\Lambda (R)$
for the corresponding void sizes $\Lambda$. We have
illustrated this dependency with regular fractals in $d=1$, 
for which exact relations
have been derived. Next, the introduction of a simple algorithm to identify
voids in any dimension has allowed us to show that the relation
$\Lambda (R) \simeq R^{-d/D}$ also holds in stochastic fractals defined in
dimension $d=2$. We have shown that the mean size of voids in a sample
defined between a lower and a higher cut-off scales with
these quantities, ${\bar \Lambda} \propto \Lambda_u^{\beta_u} 
\Lambda_l^{\beta_l}$. The exponents $\beta_u$ and $\beta_l$ depend on
the fractal dimension $D$; hence, the relation between $\bar \Lambda$ and the
two cut-offs depends on the fractal.
Our results can be straightforwardly extrapolated to $d=3$.

This study has been performed with the aim of applying it to current measures
of the distribution of galaxies. On the one hand, current measures of the
two-point correlation function seem to be consistent with a fractal 
distribution, with a yet uncertain dimension $1<D
< 2$, and in a still controversial range from $1$ to, perhaps, $\sim 100$ h$^{-1}$ Mpc (or even more) (Guzzo, 1997; Sylos Labini et al., 1998; Mart\'{\i}nez,
1999). 
On the other hand, current void catalogues (Kauffmann \& Fairall, 1991; 
Hoyle \& Vogeley, 2002) do not seem to support
this result. Nonetheless, attending the discussion which conforms the
body of our work, they are insufficient to discard
the hypothesis of a fractal distribution of galaxies. 
To assess the capability of void finding
algorithms to detect fractal structure and then the fractal dimension 
$D$, we would suggest that they 
be tested with simple examples, for which exact results can be
easily obtained, as we have done here.

It would be interesting to extend the measures of void sizes to
scales smaller than the ones usually probed in a systematic way and,
specifically, compare them with a scaling distribution.  The careful
construction of the Zipf plot of void sizes and the
analysis of its scaling (or not), as well as the scaling of the
average size of voids with the sample size are complementary measures
to $n-$point correlation functions and additional support for the
fractality of the distribution of galaxies. In any case, it is clear
that the two methods must provide equivalent information, meaning
that the study of the convergence of both approaches could help
distinguish different sources of deviation from scaling and moreover better
characterize the morphology of the galaxy distribution.

\section*{Acknowledgments}

We thank Ugo Bastolla, Alvaro Dom\'{\i}nguez, Antonino Giaquinta 
and Juan~P\'erez-Mercader 
for interesting discussions and support. The work of
both authors is currently 
supported by respective Ram\'on y Cajal contracts of the
Ministerio de Ciencia y Tecnolog\'{\i}a.

\appendix

\section[]{Zipf's law for a deterministic fractal}

We show here that the function $\Lambda(R)$ has (approximately) equal steps
in logarithmic scale and
how to eliminate the discrete variable $i$ from equations
(\ref{L}) and (\ref{R}).

We can asymptotically approximate Eq.~(\ref{R}) by
\begin{equation}
\label{approx}
R_i = 1- \frac{m}{N-1} + m\frac{N^{i-1}}{N-1}\approx m \frac{N^{i-1}}{N-1}
\end{equation}
for large $R_i$ (which implies that $i$ is large,
assuming that $N$ and $m$ are not). The step length in logarithmic scale
is
\begin{equation}
\ln(R_i + m N^{i-1}) - \ln R_i = \ln\left(1 + \frac{m N^{i-1}}{R_i}\right)
\approx \ln N.
\end{equation}

On the other hand, within the same approximation, after taking logarithms
of equations (\ref{L}) and (\ref{R}),
\begin{eqnarray}
\ln \Lambda_i = \ln c + i\,\ln r, \\
\ln R_i = \ln \frac{m}{N-1} + (i-1)\,\ln N.
\end{eqnarray}
Now, it is easy to solve for $i$ in the second equation and substitute it
in the first one, obtaining
\begin{eqnarray}
\ln \Lambda_i &=& \ln r \left(\frac{\ln R_i - \ln [m/(N-1)]}{\ln N} +1
\right) + \ln c  \nonumber \\ 
&=& \frac{-1}{D} \left(\ln R_i + \ln\frac{N-1}{m} \right) + \ln r + \ln c.
\end{eqnarray}
From $c m + N = 1/r$, $r c = (1-rN)/m$. Then, after removing the logarithms,
\begin{equation}
\Lambda_i = \left(\frac{N-1}{m}\right)^{-1/D} \frac{1-rN}{m}\, R_i^{-1/D}.
\end{equation}

Let us briefly analyze the accuracy of the approximation (\ref{approx}).
For the example in Fig.~1, with $N=5$ and $m=1,2,4$, $R_3 = 7,13,25$,
whereas the approximation yields 6.25, 12.5, 25. Of course,
the accuracy is higher for $i>3$. In general, the relative error is
$O(N^{-i})$.

\bsp

\label{lastpage}

\end{document}